# Anomalous Non-Hermitian Topological Anderson Insulator

Mina Ren,[1,2] Xi Shi,[1] Haitao Jiang,[2] Feng Liu,[1,*] Hong Chen,[2] and Yong Sun[2,†]

[1] *Department of Physics, Shanghai Normal University, Shanghai 200234, China.*
[2] *MOE Key Laboratory of Advanced Micro-Structured Materials, School of Physics Science and Engineering, Tongji University, Shanghai 200092, China.*

*fliu@shnu.edu.cn
†yongsun@tongji.edu.cn

Strong disorder drives conventional Hermitian systems into Anderson insulating states, suppressing all topological phases. Here, we unveil symmetry-protected, anomalous topological phases in the strong disorder limit of a non-Hermitian system, characterized by a scale-invariant merging of zero-energy modes. Using the maximally symmetric $J_x$ lattice as an ideal platform and introducing specifically engineered (ABBA-type) symmetry-preserving non-Hermitian disorder, we observe a sequence of disorder-induced phase transitions: from a trivial insulator into and through a non-Hermitian topological Anderson insulator (TAI) phase, culminating in a stable anomalous non-Hermitian TAI phase characterized by a quantized polarization $P_x \approx 0.25$. Within this anomalous phase protected by the mobility gap, the zero-energy modes exhibit a distinct ($N/2$)-mode coalescence that scales with system size. Our findings demonstrate that non-Hermitian disorder engineered to preserve symmetry can induce and protect novel topological order inaccessible to conventional Hermitian disorder, thereby advancing the fundamental understanding of topological phenomena mediated by the interplay of disorder and non-Hermiticity.

## I. INTRODUCTION

The investigation of topological Anderson insulators (TAIs) [1-3] has fundamentally reshaped the classification of topological phases, establishing disorder as a potent dimension for inducing and manipulating nontrivial topology [4-32]. In parallel, the study of non-Hermitian systems, characterized by gain, loss, and nonreciprocal couplings, has uncovered a wealth of phenomena absent in Hermitian settings, such as exceptional points [33-36], the non-Hermitian skin effects [37-41], anomalous localization transitions [42-44], and among others [45-50]. A key frontier lies at their confluence: the synthesis of disorder physics and non-Hermitian dynamics offers a powerful paradigm to engineer novel topological matter, exemplified by the emergence of non-Hermitian TAIs [51-58].

To isolate and elucidate the distinct roles of disorder and non-Hermiticity, an ideal, maximally symmetric platform is essential. The $J_x$ lattice [59] serves this purpose perfectly. Its pristine Hamiltonian maps exactly onto the SU(2) angular momentum operator $J_x$, endowing it with a perfectly equidistant eigenvalue spectrum (see Appendix A), an intrinsic property that provides a pristine, analytically tractable baseline against which disorder-induced effects can be unambiguously identified. This unique spectral structure has already enabled breakthroughs in photonics, including perfect state transfer [60-64], broadband absorption [65,66], optical isolation [67], and beam splitters [68].

In this Letter, we employ this ideal platform to reveal how symmetry-preserving non-Hermitian disorder can drive a system through a hierarchy of topological phases inaccessible under conventional disorder. Specifically, we introduce a designed (ABBA-type) non-Hermitian disorder profile that preserves a key pseudo-anti-Hermitian symmetry [69]. We report a sequence of disorder-induced transitions in the $J_x$ lattice: from a trivial insulator ($P_x = 0$) through a conventional non-Hermitian TAI phase ($P_x = 0.5$), culminating in a stable anomalous non-Hermitian TAI phase characterized by a quantized polarization $P_x \approx 0.25$ in the strong disorder limit.

This anomalous phase defies the conventional expectation that strong disorder universally localizes all states and destroys topology. Instead, within the mobility gap (see Appendix B), we observe the protected coexistence of topological edge states and localized bulk states. A hallmark of this phase is an unprecedented scale-invariant merging of zero-energy modes, exhibiting a distinct ($N/2$)-mode coalescence that grows with system size.

Our analysis, integrating eigenvalue spectra and real-space polarization [54], demonstrates that non-Hermiticity (see Appendix C) primarily drives the transition into the conventional TAI, while the specific symmetry-preserving non-Hermitian disorder structure is crucial for stabilizing the anomalous $P_x \approx 0.25$ phase. This work establishes that disorder engineered to preserve symmetry can induce novel topological order where generic disorder would fail, advancing the fundamental understanding of topological phenomena mediated by the interplay of non-Hermiticity and correlated disorder.

## II. MODEL AND METHODS

We consider a one-dimensional trivial $J_x$ lattice with $N$ sites, as schematically shown in Fig. 1(a). The nearest-neighbor coupling follows a parabolic profile, $J_n = \sqrt{n(N-n)}/2$ for $n = 1, 2, \cdots, N-1$, which is intrinsic to the $J_x$ lattice. This specific coupling results in a perfectly equidistant eigenvalue spectrum, placing the pristine system in a topologically trivial phase. A fundamental question we address is what novel topological physics can be induced by introducing non-Hermitian disorder into such a highly symmetric baseline.

To this end, we introduce a specifically designed, symmetry-preserving non-Hermitian disorder. As illustrated in Fig. 1(b), the disorder follows an ABBA-type pattern per unit cell: four consecutive sites are assigned alternating gain and loss, quantified by $i\gamma_s$ and $ig_s$, respectively. This pattern preserves a key pseudo-anti-Hermitian symmetry essential for topological protection [69,70]. The Hamiltonian of the disordered system is given by the tight-binding model:

$$H = \sum_{n=1}^{N-1} J_n(c_n^\dagger c_{n+1} + H.c.) \\ - \sum_{s=1}^{N/4} [ig_s(c_{4s-3}^\dagger c_{4s-3} + c_{4s}^\dagger c_{4s}) \\ + i\gamma_s(c_{4s-2}^\dagger c_{4s-2} + c_{4s-1}^\dagger c_{4s-1})] \quad (1)$$

where $c_n$ ($c_n^\dagger$) is the annihilation (creation) operator at site $n$. The gain ($ig_s$) and loss ($i\gamma_s$) strengths within the $s$-th unit cell are independent random variables, uniformly distributed in $[0, 0.5W]$ and $[-0.5W, 0]$, respectively. Here, $W$ serves as the overall disorder strength.

The topology of this one-dimensional disordered non-Hermitian lattice can be characterized by the real-space polarization $P_x$ [54], defined as

$$P_x = \frac{1}{2\pi}\text{Im}\left\{\log\left[\det(U_l^\dagger \hat{Q} U_r)\sqrt{\det(\hat{Q}^\dagger)}\right]\right\} \quad (2)$$

with $\hat{Q} = \exp(i2\pi\hat{x}/N)$. Here, $\hat{x}$ is the coordinate operator, $N$ is the total number of lattice sites, and $U_l$ ($U_r$) denotes the matrix of occupied, normalized left (right) eigenstates. This invariant robustly characterizes the topological phases across the disorder-driven transitions explored in this work.

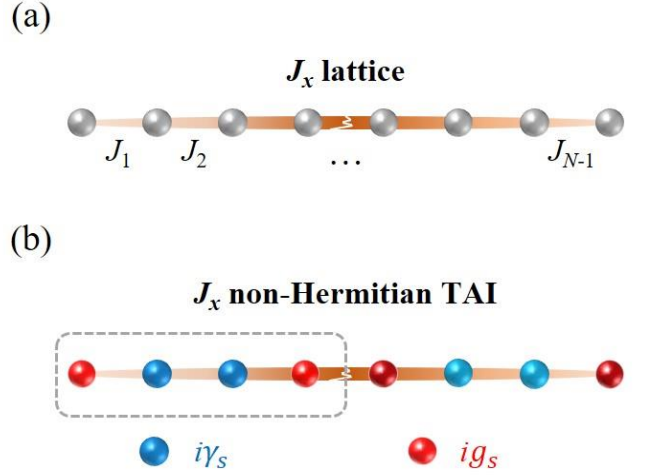

FIG. 1. $J_x$ non-Hermitian TAI. (a) Illustration of a trivial $J_x$ lattice characterized by parabolic coupling, denoted as $J_n$, with all lattice sites exhibiting identical on-site potentials. These sites are depicted as grey spheres. (b) Representation of a $J_x$ non-Hermitian TAI induced by disordered on-site potentials. Each ABBA-type unit cell comprises four sites, with on-sites imparting loss and gain, quantified by $i\gamma_s$ and $ig_s$, respectively. These are visually distinguished by blue spheres (loss) and red spheres (gain).

## III. NUMERICAL RESULTS

### A. Topological characterization of anomalous non-Hermitian TAIs

Figure 2 illustrates the evolution of the eigenspectrum and real-space polarization $P_x$ with increasing strength $W$ of the non-Hermitian disorder, revealing the intricate interplay between disorder, non-Hermiticity, and topology.

For weak disorder ($0 \lesssim W \lesssim 2.6$), the system remains in a topologically trivial phase with $P_x = 0$ [Figs. 2(a) and 2(b1)]. In this regime, the spectrum displays complex-conjugate eigenvalue pairs, a signature of protection by anti-PT symmetry. As $W$ increases, the real parts of the band-edge eigenvalues [the $N/2$-th and ($N/2+1$)-th states] begin to converge, while a pronounced sub-gap opens among the bulk modes. A slight splitting also develops in their imaginary parts [inset of Fig. 2(a2)]. In the interval $2.6 \lesssim W \lesssim 5.3$, prior to the complete merging of zero-energy modes, the polarization $P_x$ increases gradually with $W$. This smooth crossover behavior, distinct from the abrupt transition in Hermitian TAIs, results from the statistical average over disorder configurations, as the critical disorder strength $W_c$ varies across different realizations. At $W \approx 5.3$, the band-edge eigenvalues merge at zero energy, signifying the onset of a non-Hermitian TAI phase. The corresponding polarization, approximately 0.4 for this finite system, approaches the quantized value of 0.5 in larger systems, as shown in Fig. 2(b2). The inverse participation ratio (IPR) [71,72] distribution in Fig. 2(a) confirms that these zero-energy modes are significantly more localized than the bulk states.

Strikingly, under strong non-Hermitian disorder limit, the system does not evolve into a conventional Anderson insulator but instead stabilizes into a novel phase. The merged zero-energy modes persist, and adjacent eigenvalues further converge toward zero energy, with their IPR approaching unity, indicating strong localization. Most notably, as demonstrated in Fig. 2(b2) for disorder strengths up to $W = 10^4$, the polarization saturates at a quantized value of $P_x \approx 0.25$. We identify this stable phase, characterized by $P_x \approx 0.25$ and protected merged zero modes within a mobility gap, as the anomalous non-Hermitian TAI.

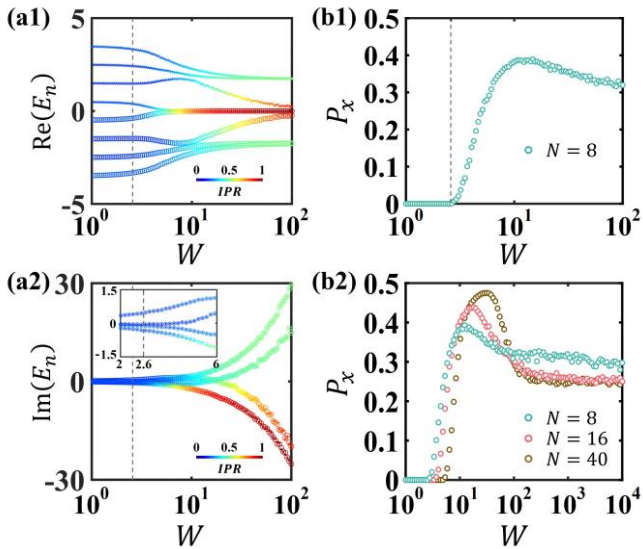

FIG. 2. Evolution of the eigenspectrum and polarizations. (a) The real part $\text{Re}(E_n)$ and the imaginary part $\text{Im}(E_n)$ of the system's eigenvalues as a function of $W$ in the interval $[10^0, 10^2]$. The disorder-averaged IPR of the eigenstates, indicative of their localization, varies with $W$ and is represented by distinct colored markers. Occupied eigenvalues ($\text{Re}(E_n) < 0$) are marked with circles, while unoccupied eigenvalues ($\text{Re}(E_n) > 0$) are denoted by pentagrams. (b) The evolution of the polarization $P_x$ with respect to $W$. The data represent an ensemble average derived from 1000 distinct disorder configurations.

Based on these results, the system exhibits three characteristic regimes as disorder strength increases: (i) a trivial insulating phase at weak disorder ($0 \lesssim W \lesssim 2.6$), (ii) an intermediate crossover regime ($2.6 \lesssim W \lesssim 5.3$) where the polarization evolves smoothly, and (iii) a topologically nontrivial phase for $W \gtrsim 5.3$. The nontrivial phase further separates into a conventional non-Hermitian TAI with quantized polarization $P_x = 0.5$ and, in the strong-disorder limit, an anomalous non-Hermitian TAI with stabilized polarization $P_x \approx 0.25$.

To further elucidate the nature of these phases, we examine the probability density distributions of key eigenstates in Fig. 3. In the topologically trivial $J_x$ lattice [Fig. 3(a)], the density profile is symmetric due to the dominant parabolic coupling. Within the crossover regime at $W = 4$ [Fig. 3(b), $P_x = 0.15$], the influence of non-Hermitian disorder becomes apparent, competing with the coupling and inducing enhanced localization at the lattice edges. In the conventional TAI phase [Fig. 3(c)], the topologically protected zero-energy modes are sharply localized at the two ends of the chain.

The anomalous TAI phase [Fig. 3(d)] exhibits a unique spatial structure: strongly localized states appear not only at the ends ($n = 1, 8$) but also at the center of the lattice ($n = 4, 5$). This reveals the coexistence of topological edge states and strongly localized bulk states at zero energy, a hallmark of the anomalous phase induced by symmetry-preserving non-Hermitian disorder. Furthermore, as detailed in Appendix C, while non-Hermiticity is the primary driver of the conventional TAI phase ($P_x = 0.5$), the stabilization of the anomalous phase ($P_x \approx 0.25$) is predominantly a consequence of the strong, correlated disorder.

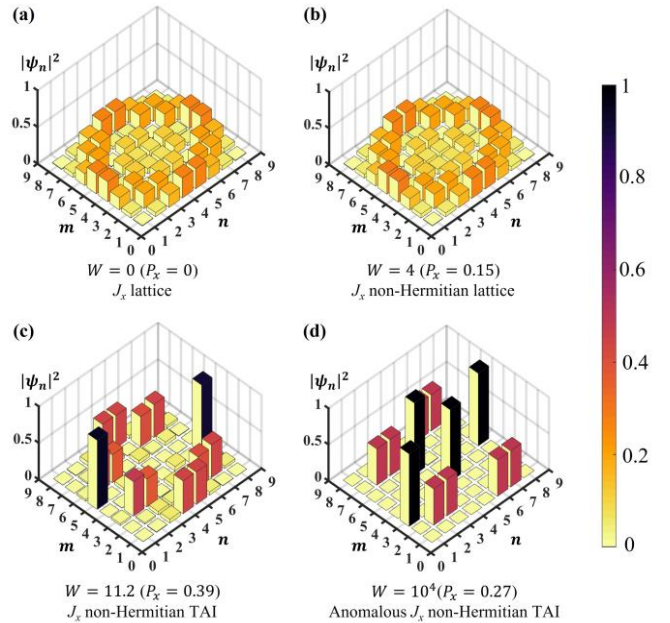

FIG. 3. Probability density distributions of eigenstates for a $J_x$ lattice of finite size $N = 8$ under varying disorder strengths: (a) initial $J_x$ lattice at $W = 0$, (b) $J_x$ non-Hermitian lattice at $W = 4$, (c) $J_x$ non-Hermitian TAI lattice at $W = 11.2$, and (d) Anomalous $J_x$ non-Hermitian TAI lattice at $W = 10^4$. The label $m$ denotes the eigenstate index, while $n$ indicates the corresponding component.

## B. Scaling merging in anomalous non-Hermitian TAIs

The hallmark of the anomalous non-Hermitian TAI phase is a characteristic merging of zero-energy modes that scales with system size. Figures 4(a1)–(c1) plot the real parts of the eigenvalues for different system sizes ($N = 8, 12, 16$) against disorder strength $W$. Under strong disorder ($W = 10^4$), a striking pattern emerges: we observe the merging of exactly $N/2$ states at zero energy (i.e., 4, 6, and 8 merged modes for the respective system sizes). This indicates that the number of coalescing zero modes grows linearly with $N$, confirming a scale-invariant merging phenomenon that is intrinsic to the anomalous phase.

To visualize the spatial structure of these merged modes, Figs. 4(a2)–(c2) display the probability density distributions at $W = 10^4$. At zero energy, the $N/2$ merged states consist of two topological edge modes sharply localized at the boundaries, accompanied by $N/2-2$ strongly localized bulk modes. This provides direct real-space evidence of the coexistence of topological edge states and localized bulk states at zero energy, a defining signature of the anomalous TAI phase induced by symmetry-preserving non-Hermitian disorder.

Together, these results establish that the anomalous TAI phase is characterized by a scale-invariant, ($N/2$)-fold coalescence of zero-energy modes, protected within the mobility gap and featuring a hybrid spatial structure of edge and bulk localization.

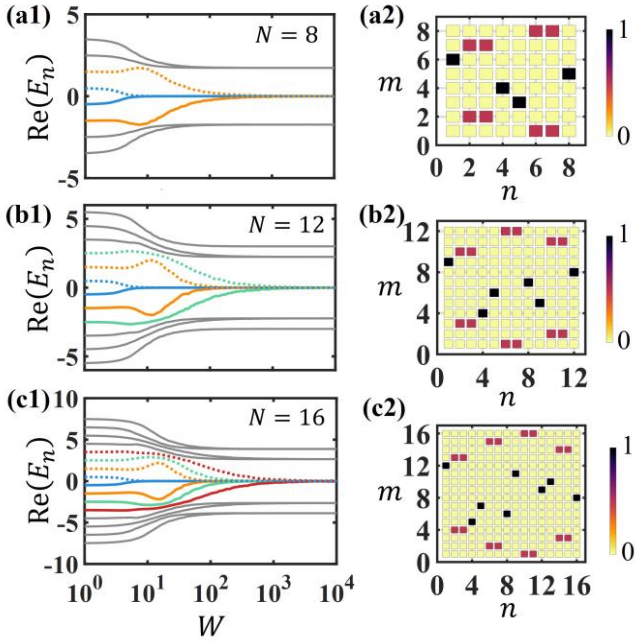

FIG. 4. Scaling merging in ultra-strong disordered non-Hermitian systems. Evolution of the real parts of eigenvalues as a function of $W$ for $J_x$ lattices of different sizes: (a1) $N = 8$, (b1) $N = 12$, and (c1) $N = 16$. The probability density distributions of eigenstates for the corresponding systems at a disorder strength of $W = 10^4$ are depicted in panels (a2-c2). The largest four occupied eigenvalues are marked with solid-colored lines, while the smallest four unoccupied eigenvalues are distinguished by dashed colored lines.

## IV. SUMMARY AND DISCUSSION

In summary, we have demonstrated that strong non-Hermitian disorder, when engineered to preserve symmetry, can induce and stabilize an anomalous TAI phase in the maximally symmetric $J_x$ lattice. This phase is characterized by two hallmark features: (i) a quantized polarization of $P_x \approx 0.25$ in the strong-disorder limit, and (ii) a scale-invariant, ($N/2$)-fold merging of zero-energy modes.

Through systematic analysis of the eigenspectrum and real-space polarization, we mapped out the disorder-driven phase evolution: the system transitions from a trivial insulator ($P_x = 0$) through a conventional non-Hermitian TAI phase ($P_x = 0.5$), finally stabilizing into the anomalous TAI phase ($P_x \approx 0.25$) at strong disorder. Crucially, this final nontrivial phase persists under extreme disorder strength rather than collapsing into a trivial Anderson insulator.

The anomalous phase is protected by a mobility gap and hosts a unique coexistence of topological edge states and strongly localized bulk states at zero energy. The number of merged modes scales precisely as $N/2$, a direct manifestation of the symmetry-preserving, correlated nature of the introduced ABBA-type disorder. This phenomenon is distinct from effects driven purely by on-site non-Hermiticity and underscores disorder correlation and symmetry as central design principles for engineering novel topological matter.

The generality of this mechanism is further demonstrated in the paradigmatic Su-Schrieffer-Heeger (SSH) model. As detailed in Appendix D, the same symmetry-preserving disorder induces an anomalous TAI phase characterized by the identical hallmarks: a quantized polarization $P_x \approx 0.25$ and a scale-invariant merging of zero-energy modes whose number grows as $N/2$.

Our findings advance the fundamental understanding of the interplay between non-Hermiticity and disorder in sculpting topological phases. They suggest that tailored disorder profiles can create topological environments where conventional Hermitian disorder would fail. As outlined in Appendix E, the predicted phenomena are directly accessible in non-Hermitian electrical circuit arrays, providing a clear path for experimental verification. This work opens avenues for harnessing disorder-engineered topological states in photonic and quantum synthetic platforms, where the coexistence of topological edge states and localized bulk states could enable new functionalities in wave control and topological lasing.

## ACKNOWLEDGMENTS

The authors acknowledge the financial support from the National Key R&D Program of China (Grant No.

2021YFA1400602 and No. 2020YFA0211402), the National Natural Science Foundation of China (NSFC; Grants No. 12141303, No. 12073018, No. 12474316 and No. 12274325). Shanghai Academic Research Leader Under Grant 22XD1422100. Shanghai Pujiang Program (21PJ1411400).

**APPENDIX A: Spectral signatures and topological properties of the conventional $J_x$ Lattice**

In the conventional $J_x$ lattice with absent non-Hermitian disorder ($W = 0$), the inter-site coupling strength $J_n = \sqrt{n(N-n)}/2$ ($n = 1,2,\cdots,N-1$) demonstrates a parabolic spatial profile, as characterized in Fig. 1(a) of the main text. Crucially, this specific coupling profile is not arbitrary but is fundamentally designed to ensure that the system's Hamiltonian becomes mathematically equivalent to the $\hat{J}_x$ angular momentum operator within the SU(2) Lie algebra. This fundamental symmetry dictates the lattice's equidistant eigenvalue spectrum, which is a direct mathematical consequence of the angular momentum quantization rules, akin to the energy-level structure of quantum harmonic oscillators, rendering it an ideal platform for high-fidelity state manipulation via spectral control [62,65,67,68].

The spectral characteristics are further elucidated through numerical simulations of a conventional 16-site $J_x$ lattice (Fig. 5). Panel 5(a) demonstrates the formation of a strictly real, equally spaced eigenvalue spectrum, which directly originates from the parabolic coupling architecture. Further analysis uncovers a parity-dependent spectral quantization mechanism in the $J_x$ lattice architecture: systems with odd site numbers ($N \equiv 1 \bmod 2$) exhibit integer eigenvalues, while even-numbered configurations ($N \equiv 0 \bmod 2$) display half-integer eigenvalues. In Fig. 5(b), the probability density distribution of the eigenstates, denoted as $|\psi_n|^2$, exhibits four prominent peaks, two of which are located at the boundaries of the $J_x$ lattice at sites $n = 1$ and $n = 16$, with the other two peaks situated near the center of the lattice at sites $n = 8$ and $n = 9$. This peculiar distribution underpins the system's capability for perfect state transfer. The $J_x$ lattice provides a perfectly understood, highly symmetric initial condition, termed the "pristine baseline", which makes the model uniquely suited for investigating disorder effects. When we introduce perturbations like our non-Hermitian disorder, they directly disrupt this perfect symmetry. The consequences manifest as distinct signatures across three fundamental aspects: systematic deviations from the ideal equidistant eigenvalue spectrum, unambiguous localization transitions in the eigenstate profiles, and emergent topological characteristics in the polarization response. Against this clean background, each of these effects can be unambiguously attributed to the specific nature of the introduced disorder.

Remarkably, our real-space polarization analysis on the conventional $J_x$ lattice, previously unexplored in this lattice, yields a zero polarization $P_x = 0$, signifying the system's trivial topological classification in the absence of disorder. This establishes the crucial starting point from which non-Hermitian disorder can induce the novel topological phases reported in our main text.

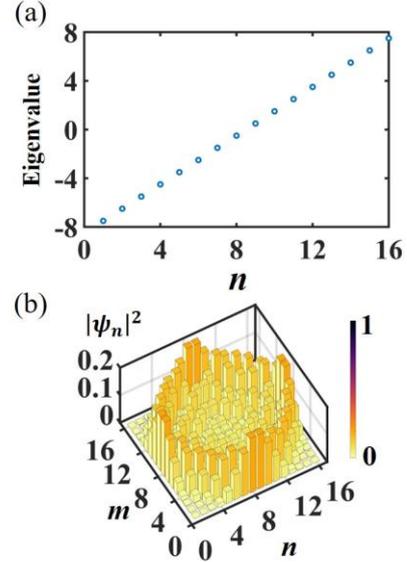

FIG. 5. Eigenspectrum and spatial distribution of eigenstates for a $J_x$ lattice with $N = 16$ sites. (a) Equidistant eigenvalue spectrum. (b) Spatial distribution of the eigenstates with index $m$.

**APPENDIX B: Arithmetic mean $\rho_{ave}$ and geometric mean $\rho_{typ}$ of the LDOS**

In Hermitian systems, edge states of topological Anderson insulators (TAIs) are protected by mobility gaps [73,74]. To elucidate the gap characteristics of two distinct non-Hermitian TAIs induced by disorder (as demonstrated in the main text), we systematically analyze the arithmetic ($\rho_{ave}$) and geometric ($\rho_{typ}$) means of the local density of states (LDOS) across topological phases under varying non-Hermitian disorder strengths (Fig. 6).

The topologically trivial Hermitian $J_x$ lattice exhibits finite $\rho_{ave}$ and $\rho_{typ}$ [Fig. 6(a)], consistent with its equidistant eigenvalue spectrum. Crucially, the gradual introduction of non-Hermitian disorder drives progressive merging of spectral edges near zero energy [Figs. 6(b)-(d)], concomitant with a continuous topological transition where polarization $P_x$ evolves from 0 to 0.5 (saturation limited by finite-size effects). This smooth evolution starkly contrasts with the abrupt jumps of topological invariants observed in Hermitian systems. In the topologically nontrivial $J_x$ non-Hermitian TAI phase [Fig. 6(d)], $\rho_{ave}$ vanishes at doubly merging zero-energy states while $\rho_{typ}$ (blue needle-like lines) persists, directly evidencing mobility gap protection. At ultra-strong non-Hermitian disorder ($W = 10^4$), complete suppression of $\rho_{ave}$ coexists with finite $\rho_{typ}$ at fourfold-merging zero energy, defining an *anomalous $J_x$ non-Hermitian TAI phase*. Importantly, both the conventional and anomalous TAI phases maintain mobility gap protection under non-Hermitian disorder. This robust topology under ultra-strong non-Hermitian disorder fundamentally contrasts with Hermitian AIs, which trivialize under ultra-strong disorder, thereby demonstrating the unique role of non-Hermitian disorder in enhancing topological robustness.

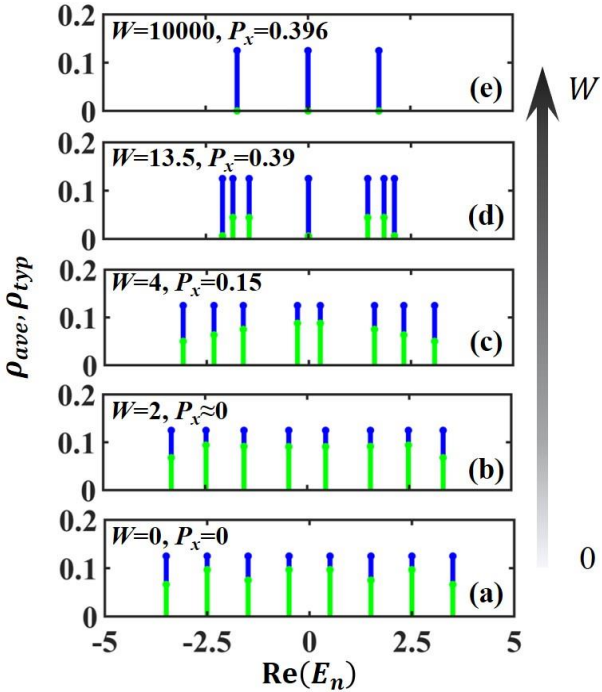

FIG. 6. Arithmetic mean $\rho_{ave}$ (blue needle-like lines) and geometric mean $\rho_{typ}$ (green needle-like lines) of the LDOS across distinct topological phases. (a) Topologically trivial Hermitian $J_x$ lattice. (b) Topologically trivial non-Hermitian $J_x$ lattice. (c) Transitional phase region of the $J_x$ non-Hermitian lattice. (d) Topologically nontrivial $J_x$ non-Hermitian TAI. (e) Topologically nontrivial anomalous $J_x$ non-Hermitian TAI. Non-Hermitian disorder strength increases vertically from bottom to top. The lattice contains $N$=8 sites.

## APPENDIX C: Engineered vs. Random Non-Hermiticity in the $J_x$ Lattice: A Comparative Study

### a. Disorder-free $J_x$ non-Hermitian lattice: spectral engineering and topological response

To systematically investigate how non-Hermiticity drives topological phase transitions in $J_x$ lattices, we first examine the system under a periodic, disorder-free ABBA-type non-Hermitian onsite potential, maintaining the fundamental lattice geometry shown in Fig. 1(b) while varying only the strength of this purely non-Hermitian modulation. Figure 7 presents the evolution under a balanced gain/loss (gain: $0.5V$, loss: $-0.5V$, where $V$ denotes the non-Hermitian strength). In Panel 7(a), the real-space polarization $P_x$ undergoes an abrupt transition from $P_x = 0$ to $P_x = 0.5$ with increasing $V$, followed by a moderate decrease at extremely strong non-Hermitian strengths. To elucidate this phenomenon, we conducted finite-size scaling analyses for systems with $N = 8$, 16, and 40 sites (represented by cyan, pink, and brown circles, respectively). The observed polarization reduction is attributed to finite-size effects, confirming that strong non-Hermiticity alone can drive a sharp topological phase transition from trivial ($P_x = 0$) to nontrivial ($P_x = 0.5$) phases.

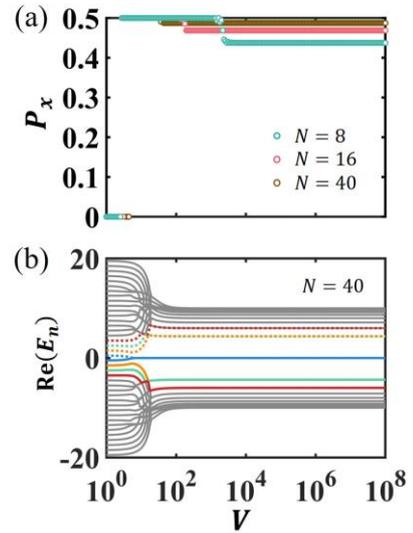

FIG. 7. For the disorder-free $J_x$ non-Hermitian lattice: (a) Evolution of real-space polarization $P_x$ as a function of the non-Hermitian strength $V \in [10^1, 10^4]$ for systems of varying sizes. Polarizations for lattices with total site numbers $N$=8, $N$=16, and $N$=40 are marked by cyan, pink, and brown circles, respectively. (b) Real parts of eigenvalues versus non-Hermitian strength $V$ for a system with $N$=40. The eigenvalues of the four dominant occupied states are highlighted with colored solid lines, while those of the four smaller unoccupied states are distinguished by colored dashed lines.

The topological phase evolution induced by pure non-Hermitian $V$ exhibits a sharp transition, contrasting sharply with the disorder-induced gradual phase transition driven by non-Hermitian disorder strength $W$ in Main Text Fig. 2(b2). This dichotomy highlights the distinct mechanisms through which non-Hermitian parameters ($V$) and disorder ($W$) govern topological reconstruction. At ultra-strong non-Hermitian disorder ($W = 10^4$), the system stabilizes in an anomalous topological phase with $P_x \approx 0.25$, distinct from the clean-system plateau at $P_x =$

0.5. This divergence highlights the critical influence of non-Hermitian disorder on topological characteristics.

Figure 7(b) reveals the $V$-dependent real eigenvalue spectrum for the $N = 40$ system, where the four dominant occupied states are plotted as solid-colored lines and the smaller unoccupied states as dashed lines. Spectral evolution demonstrates the closure of the trivial bandgap and subsequent emergence of a topological bandgap hosting doubly degenerate zero-energy modes. These spectral reconstructions provide fundamental insights into how non-Hermiticity modifies band topology, while serving as a benchmark for understanding disorder-induced spectral changes discussed in the main text.

### b. The $J_x$ lattice with generic random non-Hermitian disorder

To provide a definitive comparison with the symmetry-preserving ABBA-type disorder, we introduce random non-Hermitian disorder uniformly distributed in the range $[-i0.5W, i0.5W]$ at each lattice site, where $W$ denotes the disorder strength. This model represents a class of generic, uncorrelated non-Hermitian perturbations that lack any built-in spatial symmetry.

As shown in Fig. 8(a), the evolution of the real-space polarization $P_x$ and the inverse participation ratio (IPR) with $W$ reveals a critical distinction. While a plateau with $P_x \approx 0.25$ emerges at ultra-strong disorder ($W = 10^4$), the corresponding eigenvalue spectrum in Fig. 8(b) exhibits a complete closure of the bandgap. Furthermore, all eigenstates become strongly localized, as indicated by an IPR of 1 across the entire spectrum.

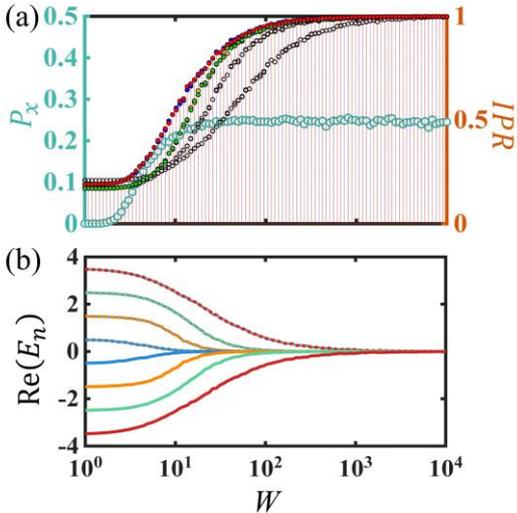

FIG. 8. $J_x$ lattice with with generic random non-Hermitian disorder. (a) Polarization $P_x$ (green circles) and disorder-averaged IPR (black circles) versus random non-Hermitian disorder strength $W$. (b) Real parts of the eigenvalues, $\text{Re}(E_n)$, as a function of $W$. All data represent an ensemble average over 1000 distinct disorder configurations.

This behavior stands in sharp contrast to the anomalous TAI phase induced by ABBA-type disorder reported in the main text. The latter phase is characterized by a robust bandgap and the emergence of protected zero-energy modes within it. The results presented here demonstrate that while generic random disorder can lead to a form of quantized polarization, it simultaneously destroys the bandgap and fails to produce the hallmark topological edge states. This confirms that the anomalous TAI phase is not a generic outcome of strong non-Hermitian disorder, but is specifically enabled by the symmetry-preserving structure of the ABBA profile, which maintains the pseudo-anti-Hermitian symmetry essential for topological protection.

### APPENDIX D: Anomalous Non-Hermitian Topological Anderson Insulator in the Su-Schrieffer-Heeger Model

Although numerical and experimental evidence for the anomalous polarization plateau ($P_x \approx 0.25$) has been reported in earlier works [57,58], this distinctive phase has not been systematically investigated. Previous studies primarily focused on the regime of moderate disorder, where conventional non-Hermitian TAIs with $P_x = 0.5$ emerge. In contrast, using a $J_x$ lattice with ideal symmetry, our work reveals that ultra-strong non-Hermitian disorder stabilizes a fundamentally distinct and previously overlooked topological phase. This phase is characterized by anomalously quantized polarization ($P_x \approx 0.25$) and a persistent, distinctive $N/2$-mode scaling merging in the ultra-strong non-Hermitian disorder limit. Here, we demonstrate that this anomalous non-Hermitian TAI phase also manifests in the Su-Schrieffer–Heeger (SSH) lattice, indicating that its formation is not constrained by specific lattice geometries, but is instead primarily governed by the introduction of the specific ABBA-type non-Hermitian disorder.

Numerical simulations are performed to determine if the introduction of ABBA-type non-Hermitian disorder into the SSH lattice leads to an anomalously quantized polarization and $N/2$-fold merging zero-energy modes. As shown in the top panel of the inset to Fig. 9(a), the SSH lattice comprises identical A and B sites (gray spheres), with intracell ($t_0$) and intercell ($t_0 = 1$) couplings distinguished by orange bonds of varying thickness. The bottom panel of the inset to Fig. 9a depicts the modified lattice structure after introducing ABBA-type on-site non-Hermitian disorder. This implementation follows the same methodology applied to the $J_x$ lattice in the main text, where each ABBA-type supercell consists of four sites with alternating gain ($ig_s$) and loss ($i\gamma_s$), represented by red and blue spheres respectively. The Hamiltonian for the disordered non-Hermitian SSH lattice, comprising $N$ lattice sites, can be described as:

$$H = \sum_{l=1}^{N/2}(t_0 c_l^\dagger c_l + t_1 c_{l+1}^\dagger c_l + H.c.)$$
$$- \sum_{s=1}^{N/4}[ig_s(c_{4s-3}^\dagger c_{4s-3} + c_{4s}^\dagger c_{4s})$$
$$+ i\gamma_s(c_{4s-2}^\dagger c_{4s-2} + c_{4s-1}^\dagger c_{4s-1})] \quad (3)$$

where $c_l$ ($c_l^\dagger$) is the annihilation (creation) operator on the $l$-th lattice cell. Within the $s$-th unit cell, we introduce additional loss and gain terms, $i\gamma_s$ and $ig_s$ ($s = 1, \cdots, N/4$), which are uniformly distributed across the intervals $[-0.5W, 0]$ for $i\gamma_s$ and $[0, 0.5W]$ for $ig_s$, with $W$ denoting the disorder strength.

For a 20-unit-cell ($N = 40$) SSH lattice, Figure 9a presents the topological phase diagram of the real-space polarization $P_x$ as a function of the intracell coupling strength $t_0$ and the non-Hermitian disorder strength $W$. It is evident that non-Hermitian disorder induces the formation of non-Hermitian TAI plateau. Notably, under ultra-strong disorder (e.g., $W = 10^4$), the real-space polarization converges to $P_x \approx 0.25$. This state is distinct from both the conventional topologically non-trivial phase ($P_x = 0.5$) and the topologically trivial AI phase found in Hermitian systems at strong disorder ($P_x = 0$). The observed anomalous polarization quantization provides further evidence that AABA-type non-Hermitian disorder serves as the dominant mechanism inducing the anomalous non-Hermitian TAI phase reported in the main text.

The eigenstate properties in the disordered non-Hermitian SSH lattice are examined using intracell coupling $t_0 = 1.05$ as an illustrative case [marked by white dashed line in Fig. 9(a)]. Figure 9b shows the evolution of the polarization and the IPR with the disorder strength $W$. The polarization values (marked by green circles) correspond to the left axis, while the IPR values (marked by black circles) correspond to the right axis. The IPRs of the eigenstates corresponding to the ($N/2$)-th and ($N/2$+1)-th eigenvalues are specifically highlighted by red and blue solid circles with black edges, respectively. With increasing non-Hermitian disorder strength $W$, the IPRs of the $N/2$ modes that ultimately merging exhibit a smooth, monotonic increase. At $W = 10^4$, the IPRs for the band-edge states (20th and 21st eigenvalues) approach unity, indicating strong localization. Surprisingly, the IPRs for a subset of bulk states also increase significantly towards unity. Figures 9(c) and 9(d) display the real and imaginary parts of the eigenvalues, respectively, as a function of $W$. These plots clearly reveal the emergence of a $N/2$-fold merging zero-energy mode under ultra-strong non-Hermitian disorder, with all associated modes exhibiting strong localization.

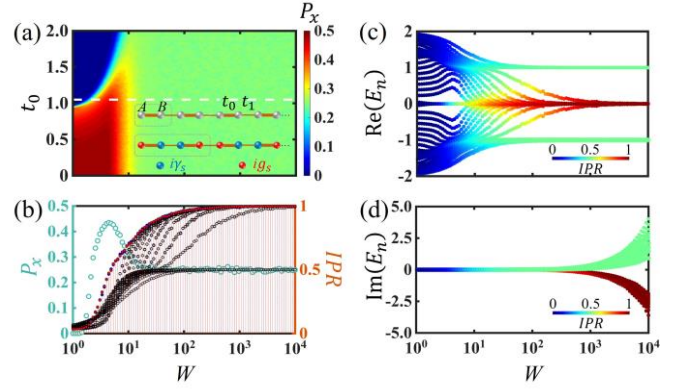

FIG. 9. Anomalous Non-Hermitian TAIs induced by ABBA-type disorder in a 20-unit-cell SSH chain. (a) Topological phase diagram of the real-space polarization $P_x$ as a function of non-Hermitian disorder strength $W$ and intracell coupling $t_0$. The insets present schematic of the lattice structures for the SSH lattice and the system with incorporated ABBA-type non-Hermitian disorder. (b) Evolution of the polarization $P_x$ (green circles) and disorder-averaged IPR (black circles) of the eigenstates with respect to $W$. (c) The real part $\text{Re}(E_n)$ and (d) the imaginary part $\text{Im}(E_n)$ of the SSH lattice's eigenvalues as a function of $W$ in the interval $[10^0, 10^4]$. All data represent an ensemble average over 1000 distinct disorder configurations.

To gain further insight into the properties of the various topological phases induced by different disorder strengths, we computed the spatial probability distributions of representative eigenstates. Figure 10 showcases these distributions for four distinct topological phases. In the SSH lattice ($W = 0$, Fig. 10a), all eigenstates are extended bulk states distributed across the entire chain. As the non-Hermitian disorder strength increases into the transition regime ($W = 2$, Fig. 10b), the spatial profiles of the eigenstates begin to exhibit enhanced localization. In conventional SSH non-Hermitian TAIs ($W = 4$, Fig. 10c), the topological edge states corresponding to the doubly merging zero-energy modes are prominently localized at the two ends ($n = 0$ and $n = 40$) of the SSH lattice. Upon further increasing the disorder strength into the stable anomalous phase where the polarization quantizes at $P_x \approx 0.25$ ($W = 10^4$, Fig. 10d), the system enters the anomalous non-Hermitian TAI phase. This phase is defined by the emergence of $N/2$-fold merging zero-energy modes, all of which are strongly localized (IPR $\approx$ 1). The remaining $N/2$ eigenstates form a set of localized bulk states with IPR $\approx 0.5$. Thus, this anomalous TAI phase exhibits the intriguing coexistence of topological edge states and two distinct types of localized bulk states.

In summary, the emergence of the anomalous TAI phase induced by ABBA-type non-Hermitian disorder is not restricted to the $J_x$ lattice. Rather, it stems from the intrinsic symmetry and topological interaction mechanisms inherent in this engineered non-Hermitian disorder configuration, which collectively give rise to such unconventional quantum phases.

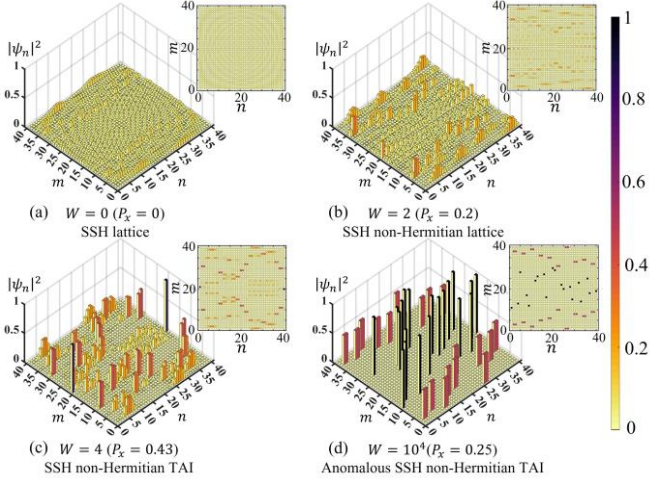

FIG. 10. Probability density distributions of eigenstates for a 20-unit-cell SSH lattice under varying disorder strengths: (a) SSH lattice at $W = 0$, (b) SSH non-Hermitian lattice at $W = 2$, (c) SSH non-Hermitian TAI lattice at $W = 4$, and (d) Anomalous SSH non-Hermitian TAI lattice at $W = 10^4$. The label $m$ denotes the eigenstate index, while $n$ indicates the corresponding component. The inset in the top-right panel provides a top-view visualization of the density distribution.

## APPENDIX E: Circuit Implementation of $J_x$ Non-Hermitian TAIs

This section presents the experimental implementation scheme of one-dimensional $J_x$ non-Hermitian TAIs through engineered circuit architectures. Figure 11(a) schematically illustrates the $J_x$ non-Hermitian lattice featuring on-site disordered gain ($ig_s$), loss ($i\gamma_s$), and parabolic coupling $J_n$. Figure 11(c) details the circuit implementation of the ABBA-type four-site non-Hermitian unit cell, where gain $ig_s$ and loss $i\gamma_s$ are physically encoded through negative impedance converter (NIC) modules ($-R_{s1}$) and positive resistors ($R_{s2}$), respectively, achieving precise alignment with theoretical predictions [Fig. 11(b)].

Each resonant site [Fig. 11(c)] comprises a parallel $L_0$-$C_0$ tank circuit establishing the fundamental resonance frequency. Parabolic coupling between nodes is engineered via capacitors $C_n$ and $C_{n-1}$, while open boundary conditions are enforced by terminating the chain with single capacitors $C_1$ and $C_{N-1}$. Trimming resistors $R_0$ suppress the imaginary components of edge-state eigenfrequencies, sharpening boundary resonance peaks. Central to gain engineering is the NIC circuit [75,76] shown in Fig. 11(b). Utilizing an operational amplifier with negative feedback, this module inverts the phase relationship between input current $I_{in}$ and voltage $V_0$ under virtual open/short conditions, generating effective negative resistance. This non-dissipative mechanism enables programmable energy injection, a critical capability for stabilizing disorder-dependent gain profiles.

The proposed feasible circuit-based protocol realizes $J_x$ non-Hermitian TAIs, building on established circuit techniques for implementing Hermitian TAIs [20] and tunable NICs that emulate non-Hermitian gain [75-77]. The circuit-based $J_x$ non-Hermitian TAI can be implemented by tuning the parameters $L_0$, $C_0$, $C_n$, $-R_{s1}$ and $R_{s2}$. This platform enables direct electrical characterization of the system's localization properties and observation of anomalous scaling merging phenomena in the $J_x$ non-Hermitian TAI phase.

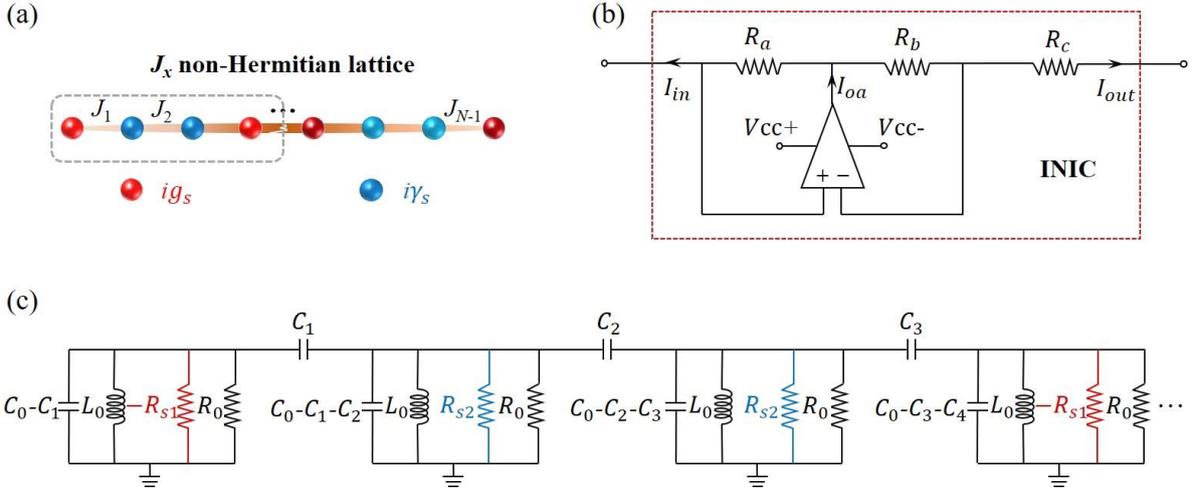

FIG. 11. Circuit implementation of $J_x$ non-Hermitian TAIs. (a) Schematic of the $J_x$ non-Hermitian lattice with on-site disordered gain $ig_s$ (red spheres) and loss $i\gamma_s$ (blue spheres). The dashed rectangle highlights a non-Hermitian unit cell of the ABBA type comprising four lattice sites. (b) Schematic diagram of the negative-resistance module $-R_{s1}$ used to implement gain $ig_s$. (c) Circuit realization of a non-Hermitian unit cell. Parabolic coupling strengths $J_n$ are achieved via lumped coupling capacitors $C_n$. The resonance frequency at each site is controlled by a parallel $LC$ resonant circuit (inductor $L_0$, capacitor $C_0$) with a trimming resistor $R_0$. On-site disordered loss $i\gamma_s$ is implemented using positive resistors $R_{s2}$, while disordered gain $ig_s$ is realized via negative-resistance modules $-R_{s1}$.